\documentclass[pdflatex,sn-aps]{sn-jnl}

\usepackage{graphicx} 
\usepackage{multirow}
\usepackage{amsmath, amssymb, amsfonts}
\usepackage{amsthm}
\usepackage{mathrsfs}
\usepackage[title]{appendix}
\usepackage{xcolor} 
\usepackage{textcomp}
\usepackage{manyfoot}
\usepackage{booktabs}
\usepackage{algorithm}
\usepackage{algorithmicx}
\usepackage{algpseudocode}
\usepackage{listings}
\usepackage{dcolumn} 
\usepackage{bm} 
\usepackage{hyperref} 
\hypersetup{
  citecolor=blue,
  colorlinks = true,
  urlcolor = blue
}
\usepackage[normalem]{ulem} 
\usepackage{physics}
\usepackage{physics2}
\usephysicsmodule{ab, ab.braket}
\usepackage{cancel} 
\usepackage{caption}
\usepackage{orcidlink}

\theoremstyle{thmstyleone}%
\theoremstyle{thmstyletwo}%

\theoremstyle{thmstylethree}%

\begin{document}

\title[Article Title]{Cusped Electrical Conductivity in Spin-1 Chiral Fermion Systems Arising from Multifold Band Degeneracy}


\author*[1]{\fnm{Risako} \sur{Kikuchi} \orcidlink{0009-0001-8649-7614}}\email{kikuchi.risako.i4@s.mail.nagoya-u.ac.jp}

\author[2]{\fnm{Junya} \sur{Endo} \orcidlink{0000-0002-2775-4780}}\email{endo@hosi.phys.s.u-tokyo.ac.jp}
\equalcont{These authors contributed equally to this work.}

\author[1]{\fnm{Ai} \sur{Yamakage} \orcidlink{0000-0003-4052-774X}}\email{yamakage.ai.x6@f.mail.nagoya-u.ac.jp}
\equalcont{These authors contributed equally to this work.}

\affil*[1]{\orgdiv{Department of Physics}, \orgname{Nagoya University}, \orgaddress{\street{Furo-cho, Chikusa-ku}, \city{Nagoya}, \postcode{4648602}, \state{Aichi}, \country{Japan}}}

\affil[2]{\orgdiv{Department of Physics}, \orgname{University of Tokyo}, \orgaddress{\street{Hongo}, \city{Bunkyo}, \postcode{1130033}, \state{Tokyo}, \country{Japan}}}


\abstract{The energy-dependent electrical conductivity in spin-1 chiral fermion systems with disorder is studied using the self-consistent Born approximation. A distinct cusp-like feature appears at an energy different from the band-crossing point, arising from the multifold band-crossing structure formed by the Dirac and trivial bands. 
By decomposing the conductivity into band-resolved intraband and interband contributions, we clarify the microscopic origin of the cusp structure previously identified in spin-1 chiral fermion systems.
These findings demonstrate the critical role of multifold band crossings and disorder-induced broadening of energy levels in determining the transport properties, offering theoretical insight into the unconventional conductivity behavior observed in topological semimetals hosting spin-1 chiral fermions.}




\maketitle

\section{Introduction}\label{sec1}

Topological semimetals are characterized by crossings of multiple energy bands, known as multifold fermions, and exhibit unique physical properties~\cite{Ma2012, Beyond2016, Lv2021}. 
A minimal model for a multifold fermion is the spin-1/2 Dirac or Weyl fermion, which arises from the crossing of two linear bands. Another notable case is the threefold band degeneracy, consisting of two linear bands and one nearly flat (or ``trivial") band, referred to as a spin-1 chiral fermion. Here, the term ``spin'' refers to an effective pseudospin degree of freedom associated with the multiband structure, not to the relativistic electron spin.
Three-dimensional spin-1 chiral fermion systems have been theoretically predicted and experimentally observed in chiral crystals~\cite{Tang2017-kk, Chang2017, Pshenay-Severin_2018, Takane2019, Rao2019-ts, Sanchez2019-by, Mozaffari2020, Robredo_2024}.

Because of their nontrivial topology and characteristic multiband structure, spin-1 chiral fermion systems have been studied in a wide range of contexts, including transport properties~\cite{Tang21, Kikuchi2022, Lien2023, Kikuchi2023, Selma2024, Haidar2025, Mandal2026}, optical response~\cite{Flicker18, Sanchez-Martinez2019-ek, Li2019, Habe2019-ss, Maulana2020, Chang2020, Dylan2020, Xu2020-zb, Le2020, Ni2020-ze, Ni2021-eo, Rees2021, Kaushik2021-hx, Dey2022-th, Lu2022, Hsiu2023, Yang2023}, and quantum phenomena associated with topological structures~\cite{Yuan2019-hz, Huber2022, hsu2022disorder, Balduini2024}. 
In particular, when examining impurity effects on electrical transport, the complex energy dependence of the multiband structure suggests that scattering processes should be treated beyond a simple constant-relaxation-time picture. 
Indeed,
the electrical conductivity behavior near the Dirac point is particularly sensitive to disorder, making a theoretical framework that accurately captures scattering effects essential~\cite{Mucciolo2010, Vigh2013, Syzranov2018, Kikuchi2022, hsu2022disorder,  Kikuchi2023, Leeb2023, Kikuchi2025}. The self-consistent Born approximation (SCBA) is a well-established method for this purpose, as it can properly incorporate the disorder-induced broadening of the density of states.
Of particular interest is the cusp-like structure in the energy dependence of the electrical conductivity, which originates from the multifold nature of the band crossings and is well described theoretically by employing the SCBA~\cite{Shon1998-ke, Noro2010-cm, 0minato2014, Ominato2015-um, Kikuchi2025}. 

For spin-1 chiral fermions, this cusp structure appears asymmetrically with respect to energy in the electrical conductivity~\cite{Kikuchi2025}. Such asymmetric cusps in the conductivity are highly intriguing, as they may lead to unconventional thermoelectric responses, which are both theoretically and practically significant.
Building on previous studies of the energy dependence of the electrical conductivity in spin-1 chiral fermion systems~\cite{Kikuchi2022,Kikuchi2025}, we focus here on how the band-resolved intraband and interband contributions depend on impurity scattering strength and the curvature of the trivial band. By systematically analyzing these dependences, we clarify how their interplay gives rise to the cusp structure and thereby influences transport responses sensitive to the detailed conductivity spectrum.

Specifically, we analyze the electrical conductivity of spin-1 chiral fermion systems with disordered potentials using the self-consistent Born approximation (SCBA) combined with vertex corrections. We show that the energy at which the cusp structure emerges, $\epsilon_c$, is determined by the strength of impurity scattering. We find that as the scattering becomes stronger, both $|\epsilon_c|$ and the corresponding conductivity $\sigma_c$ increase. Our analysis also reveals that a smaller curvature of the trivial band makes the system more susceptible to scattering, which further enhances the values of $|\epsilon_c|$ and $\sigma_c$. These results provide fundamental insights into the transport anomalies originating from multifold band crossings.
Although our focus here is on impurity-induced structures in the energy-dependent conductivity rather than on geometric quantities such as the Berry curvature or the associated anomalous transport responses, the underlying threefold band degeneracy is itself rooted in a symmetry-protected multiband crossing, and the resulting coexistence of Dirac and trivial bands plays an essential role in the formation of the cusp structure studied here.

\section{Model for a Spin-1 Chiral Fermion}\label{sec2}

We study a three-dimensional spin-1 chiral fermion with rotational and time-reversal symmetries.  
The system is described by the following effective Hamiltonian~\cite{Mandal2021-ic}:
\begin{align}
\hat{\mathcal{H}_0} = \hbar v\, \hat{\bm{S}} \cdot \bm{k} + c'\left[(\hat{\bm{S}} \cdot \bm{k})^2 - \,k^2 \hat{S}_0\right], \label{Hamiltonian}
\end{align}
where \( \bm{k} \) denotes the wave vector, and \( v \) is the Fermi velocity.  
\( \hat{\bm{S}} = (\hat{S}_x, \hat{S}_y, \hat{S}_z) \) are given in matrix form as follows:
\begin{align}
\hat{S}_{x} =
\begin{pmatrix}
0 & i & 0 \\
-i & 0 & 0 \\
0 & 0 & 0
\end{pmatrix},
\quad
\hat{S}_{y} =
\begin{pmatrix}
0 & 0 & -i \\
0 & 0 & 0 \\
i & 0 & 0
\end{pmatrix},
\quad
\hat{S}_{z} =
\begin{pmatrix}
0 & 0 & 0 \\
0 & 0 & i \\
0 & -i & 0
\end{pmatrix},
\end{align}
and \( \hat{S}_0 \) denotes the identity matrix. The second term in Eq.~\ref{Hamiltonian}, characterized by the parameter $c'$, introduces a quadratic correction to the dispersion. Specifically, this term adds curvature exclusively to the trivial band, leaving the Dirac bands linear. The parameter $c'$ determines the strength of this curvature.

\begin{figure}
  \centering
  \includegraphics[width=0.5\linewidth]{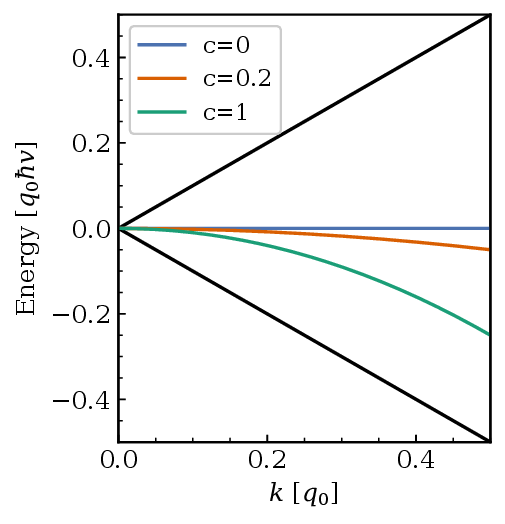}
  \caption{(Color online) Band structure of the spin-1 chiral fermion system. 
  The black solid lines denote the Dirac bands, while the colored curves represent the trivial band for different values of the curvature parameter $c$: $c=0$ (blue), $c=0.2$ (orange), and $c=1$ (green). 
  }
\label{fig_1}
\end{figure}

The energy spectrum consists of three bands, with eigenvalues given by
\begin{align}
\epsilon_{\text{c},\bm{k}} &= \hbar v k, \\
\epsilon_{\text{t},\bm{k}} &= -c' k^2, \\
\epsilon_{\text{v},\bm{k}} &= -\hbar v k,
\end{align}
where the subscripts \( \text{c}, \text{t}, \text{v} \) correspond to the conduction, trivial, and valence bands, respectively.
The corresponding band structure, illustrating the effect of the curvature parameter $c$, is shown in Fig.~\ref{fig_1}. 
We also define the dimensionless parameter characterizing the curvature of the trivial band:
\begin{align}
c=\frac{c' q_0}{\hbar v},
\end{align}
where $q_0$ is the inverse of the characteristic length scale.

To incorporate disorder effects, we consider randomly distributed impurities and introduce the impurity potential
\begin{align}
\hat{V}_{\mathrm{imp}}(\bm{r}) = \sum_i U(\bm{r}-\bm{r}_i),
\end{align}
where $\bm{r}_i$ denotes the position of the $i$th impurity and $U(\bm{r})$ is the single-impurity potential, which is assumed to have a finite-range Gaussian form.  
The total Hamiltonian is then given by
\begin{align}
\hat{\mathcal{H}} = \hat{\mathcal{H}}_0 + \hat{V}_{\mathrm{imp}}(\bm{r}).
\end{align}
 
The real-space form of the single-impurity potential is given by
\begin{align}
U(\bm{r}) = \frac{\pm u_0}{(\sqrt{\pi} d_0)^3} \exp\left(-\frac{r^2}{d_0^2}\right),
\label{gauss}
\end{align}
where \( d_0 \) characterizes the range of the potential, and \( \pm u_0 \) represents its strength.  
The sign \( \pm \) indicates that the system contains an equal number of positive and negative impurities, ensuring that the Fermi level remains unaffected by the impurity concentration.
The Fourier transform of the potential is computed as
\begin{align}
u(\bm{k}) = \int d^3 r \, e^{-i \bm{k} \cdot \bm{r}} U(\bm{r}) = \pm u_0 \exp\left(-\frac{k^2}{q_0^2}\right),
\label{u_Gauss}
\end{align}
where the inverse length scale is defined as \( q_0 = 2/d_0 \).
The scattering strength is quantified by the parameter $W$, given by
\begin{align}
W = \frac{q_0n_{\text{i}}u_0^2}{\hbar^2 v^2}.
\end{align}
Here, $n_{\text{i}}$ denotes the number of scatterers per unit volume.

\section{Self-Consistent Born Approximation}\label{sec3}
Assuming that impurities are randomly and uniformly distributed, we consider the disorder-averaged Green's function
\begin{align}
\hat{G}(\bm{k}, \epsilon + is0) 
=\left\langle\left[\epsilon \hat{S}_0-\hat{\mathcal{H}}\right]^{-1}\right\rangle
=\ab[\epsilon \hat{S}_0 - \hat{\mathcal{H}}_0 - \hat{\Sigma}(\bm{k}, \epsilon + is0)]^{-1},
\label{green function}
\end{align}
where $\langle \cdots \rangle$ denotes the configurational average over impurity positions.
The index \( s = \pm 1 \) distinguishes between the retarded (\( s=1 \)) and advanced (\( s=-1 \)) Green's functions.

The self-energy \( \hat{\Sigma} \) is determined self-consistently via the integral equation
\begin{align}
\hat{\Sigma}(\bm{k}, \epsilon + is0) = 
\int \frac{d\bm{k}'}{(2\pi)^3} n_{\text{i}} |u(\bm{k} - \bm{k}')|^2 \hat{G}(\bm{k}', \epsilon + is0).
\label{self energy}
\end{align}

The conductivity \( \sigma(\epsilon) \) at zero temperature, calculated using the Kubo formalism, is given by
\begin{align}
\sigma(\epsilon) &= 
- \frac{\hbar e^2 v^2}{4\pi} 
\sum_{s, s' = \pm 1} ss' 
\int \frac{d\bm{k}'}{(2\pi)^3} 
\,
\notag\\&\quad\times
\text{Tr}\left[
\frac{\hat{v}_x}{v} \,
\hat{G}(\bm{k}', \epsilon + is0) \,
\hat{J}_x(\bm{k}', \epsilon + is0, \epsilon + is'0) \,
\hat{G}(\bm{k}', \epsilon + is'0)
\right],
\label{conductivity}
\end{align} 
where the velocity operator in the \( x \)-direction is defined as
\begin{align}
\hat{v}_x 
&= \frac{1}{\hbar} \frac{\partial \hat{\mathcal{H}}_0}{\partial k_x} .
\end{align}

The current vertex function \( \hat{J}_x \), incorporating vertex corrections, satisfies the Bethe--Salpeter equation:
\begin{align}
\hat{J}_x(\bm{k}, \epsilon, \epsilon') 
&= \frac{\hat{v}_x}{v} + 
\int \frac{d\bm{k}'}{(2\pi)^3} 
n_{\text{i}} |u(\bm{k} - \bm{k}')|^2 
\hat{G}(\bm{k}', \epsilon) 
\hat{J}_x(\bm{k}', \epsilon, \epsilon') 
\hat{G}(\bm{k}', \epsilon').
\label{Bethe}
\end{align}

To clarify the physical origins of various contributions in a multiband system, we transform the Green's function into the band basis using the unitary matrix $\hat U$ that diagonalizes the clean Hamiltonian, $\hat{U}^\dagger \hat{\mathcal H}_0 \hat{U} = \operatorname{diag}(\epsilon_{\mathrm c, \boldsymbol k}, \epsilon_{\mathrm t, \boldsymbol k}, \epsilon_{\mathrm v, \boldsymbol k})$. 
The explicit form of $\hat U$ is given by
\begin{align}
\hat{U}
&=
\frac{1}{\sqrt{2}k\sqrt{k_x^2+k_z^2}}
\begin{pmatrix}
k_yk_z-ikk_x & \sqrt{2}\sqrt{k_x^2+k_z^2}\,k_z & k_yk_z+ikk_x\\
-k_x^2-k_z^2 & \sqrt{2}\sqrt{k_x^2+k_z^2}\,k_y & -k_x^2-k_z^2\\
k_xk_y+ikk_z & \sqrt{2}\sqrt{k_x^2+k_z^2}\,k_x & k_xk_y-ikk_z
\end{pmatrix},
\label{eq:unitary_matrix}
\end{align}
where $k = |\bm{k}|$ and $(k_x,k_y,k_z)=k(\sin\theta\cos\phi,\sin\theta\sin\phi,\cos\theta)$.
The Green's function in the band basis is then written as
\begin{align}
\hat{U}^{\dagger} \hat{G}(\bm{k}, \epsilon + is0) \hat{U} = 
\begin{pmatrix}
G^s_{\mathrm{c}} & 0 & 0 \\
0 & G^s_{\mathrm{t}} & 0 \\
0 & 0 & G^s_{\mathrm{v}}
\end{pmatrix},
\end{align}
where the labels $\mathrm{c}$, $\mathrm{t}$, and $\mathrm{v}$ correspond to the conduction, trivial, and valence bands, respectively. 
In this basis, the velocity and vertex operators are represented as:
\begin{align}
\frac{1}{v} \hat{U}^{\dagger} \hat{v}_x \hat{U} = 
\begin{pmatrix}
v_{\mathrm{cc}} & v_{\mathrm{c}\mathrm{t}} & 0 \\
v_{\mathrm{t}\mathrm{c}} & v_{\mathrm{t}\mathrm{t}} & v_{\mathrm{t}\mathrm{v}} \\
0 & v_{\mathrm{v}\mathrm{t}} & v_{\mathrm{vv}}
\end{pmatrix}, \quad
\hat{U}^{\dagger} \hat{J}_x(\bm{k}, \epsilon + is0, \epsilon + is'0) \hat{U} = 
\begin{pmatrix}
J^{ss'}_{\mathrm{cc}} & J^{ss'}_{\mathrm{c}\mathrm{t}} & 0 \\
J^{ss'}_{\mathrm{t}\mathrm{c}} & J^{ss'}_{\mathrm{t}\mathrm{t}} & J^{ss'}_{\mathrm{t}\mathrm{v}} \\
0 & J^{ss'}_{\mathrm{v}\mathrm{t}} & J^{ss'}_{\mathrm{vv}}
\end{pmatrix}.
\end{align}

The total conductivity is decomposed into physically distinct contributions as follows. 
The intraband part within the Dirac bands (conduction and valence) is given by:
\begin{align}
- \frac{\hbar e^2 v^2}{4\pi} 
\sum_{s, s' = \pm 1} ss' 
\int \frac{d\bm{k}'}{(2\pi)^3} 
\left(
v_{\mathrm{cc}} G^s_{\mathrm{c}} J^{ss'}_{\mathrm{cc}} G^{s'}_{\mathrm{c}} 
+ v_{\mathrm{vv}} G^s_{\mathrm{v}} J^{ss'}_{\mathrm{vv}} G^{s'}_{\mathrm{v}} 
\right).
\label{intra}
\end{align}

The interband term involving the Dirac and trivial bands reads:
\begin{align}
&- \frac{\hbar e^2 v^2}{4\pi} 
\sum_{s, s' = \pm 1} ss' 
\int \frac{d\bm{k}'}{(2\pi)^3}
\notag\\&\quad\times
\left(
v_{\mathrm{t}\mathrm{c}} G^s_{\mathrm{c}} J^{ss'}_{\mathrm{c}\mathrm{t}} G^{s'}_{\mathrm{t}} 
+ v_{\mathrm{c}\mathrm{t}} G^s_{\mathrm{t}} J^{ss'}_{\mathrm{t}\mathrm{c}} G^{s'}_{\mathrm{c}} 
+ v_{\mathrm{t}\mathrm{v}} G^s_{\mathrm{v}} J^{ss'}_{\mathrm{v}\mathrm{t}} G^{s'}_{\mathrm{t}} 
+ v_{\mathrm{v}\mathrm{t}} G^s_{\mathrm{t}} J^{ss'}_{\mathrm{t}\mathrm{v}} G^{s'}_{\mathrm{v}}
\right).
\label{inter}
\end{align}

Finally, the intraband contribution within the trivial band is:
\begin{align}
- \frac{\hbar e^2 v^2}{4\pi} 
\sum_{s, s' = \pm 1} ss' 
\int \frac{d\bm{k}'}{(2\pi)^3} 
v_{\mathrm{t}\mathrm{t}} G^s_{\mathrm{t}} J^{ss'}_{\mathrm{t}\mathrm{t}} G^{s'}_{\mathrm{t}}.
\label{intra_trivial}
\end{align}
Details of the derivation and numerical calculations are given in Ref.~\cite{Kikuchi2025}.

\section{Cusped Electrical Conductivity}\label{sec4}
\begin{figure}[htbp]
\centering
\captionsetup{width=0.9\textwidth} 
\includegraphics[width=0.9\textwidth]{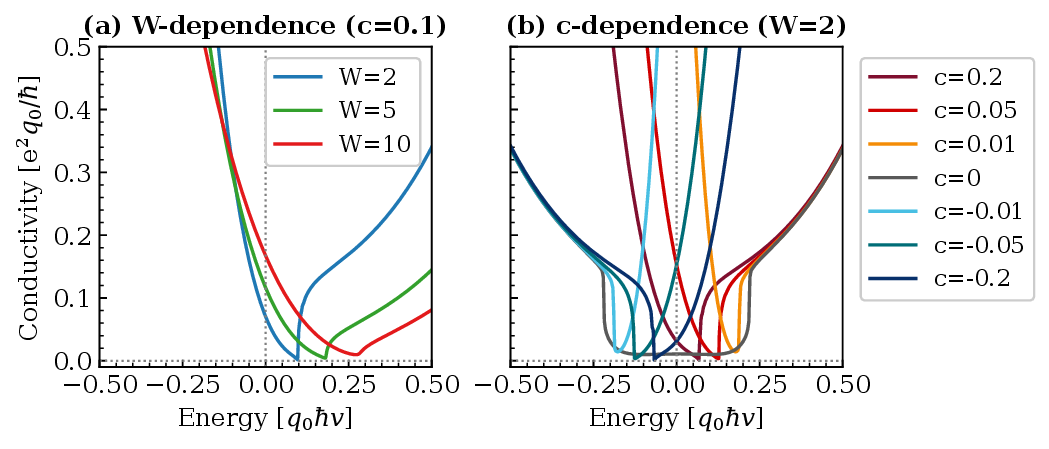}
\caption{(Color online) 
Conductivity as a function of energy for various values of the impurity parameter \( W \) and the curvature parameter \( c \).  
Panel~(a) shows the \( W \)-dependence of the conductivity spectra at a fixed curvature \( c = 0.1 \), for \( W = 2, 5, 10 \).  
Panel~(b) shows the \( c \)-dependence at a fixed impurity strength \( W = 2 \), for \( c = -0.2 \) to \( 0.2 \).  
}
\label{fig_2}
\end{figure}
\begin{figure}[htbp]
\centering
\captionsetup{width=0.9\textwidth} 
\includegraphics[width=0.9\textwidth]{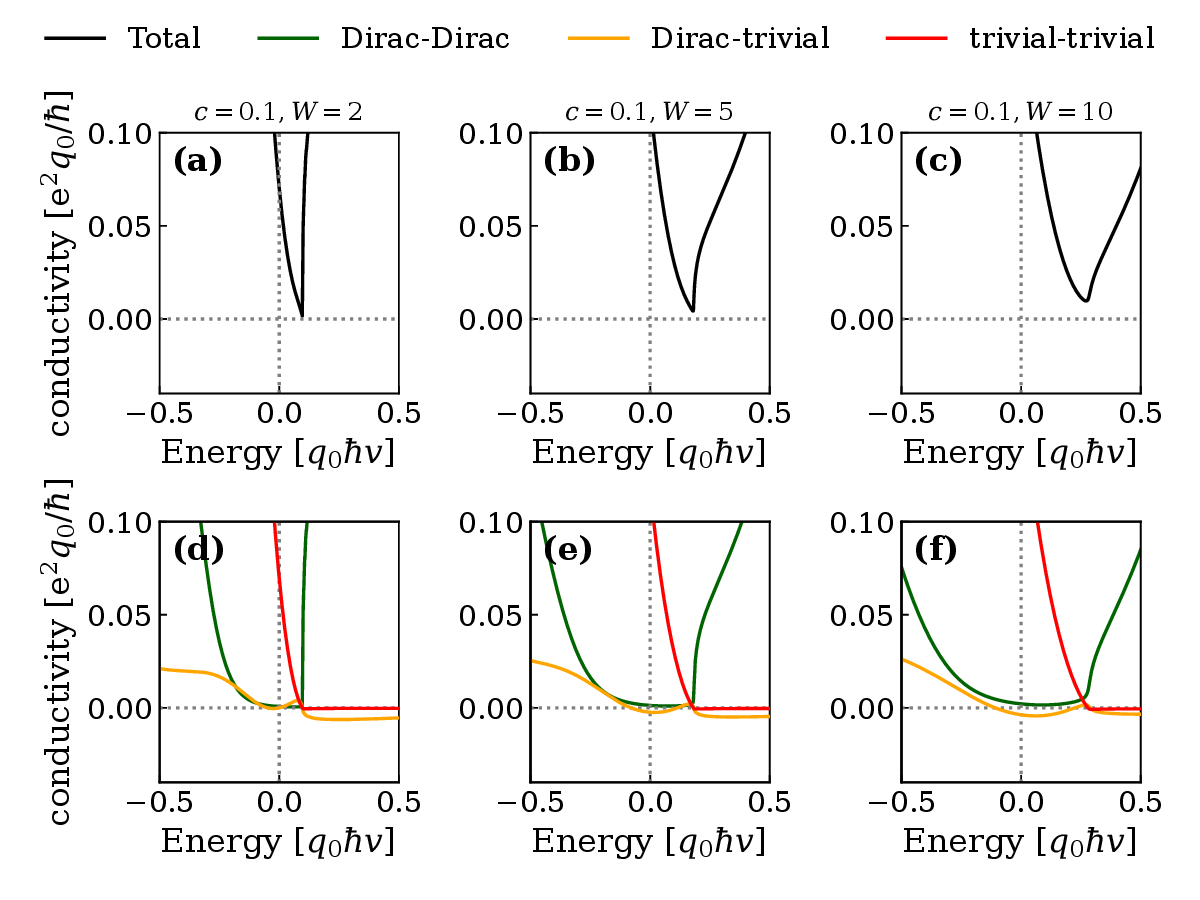}
\caption{(Color online) 
Conductivity as a function of energy for different impurity strengths \( W = 2, 5, 10 \) with fixed curvature parameter \( c = 0.1 \). 
The upper panels (a–c) show the total conductivity, while the lower panels (d-f) show the decomposed contributions: 
green lines correspond to intraband processes within Dirac bands (``Dirac--Dirac''), 
orange lines to interband processes between Dirac and trivial bands (``Dirac--trivial''), 
and red lines to intraband processes within trivial bands (``trivial--trivial''). 
}
\label{fig_3}
\end{figure}
\begin{figure}[htbp]
\centering
\captionsetup{width=0.9\textwidth} 
\includegraphics[width=0.9\textwidth]{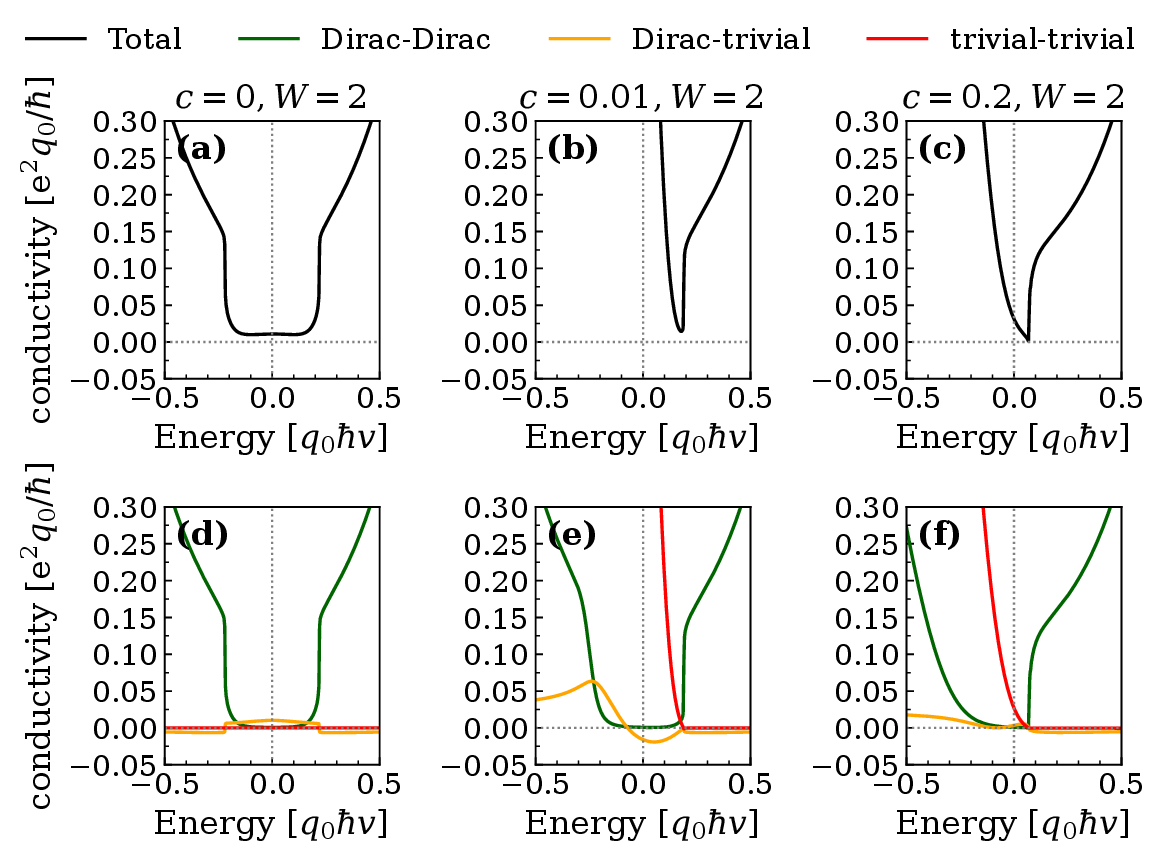}
\caption{(Color online) 
Conductivity as a function of energy for different curvatures \( c = 0, 0.01, 0.2 \) with fixed impurity strength \( W = 2 \). 
The upper panels (a-c) show the total conductivity, while the lower panels (d-f) show the decomposed contributions: 
green lines correspond to intraband processes within Dirac bands (``Dirac--Dirac''), 
orange lines to interband processes between Dirac and trivial bands (``Dirac--trivial''), 
and red lines to intraband processes within trivial bands (``trivial--trivial''). 
}
\label{fig_4}
\end{figure}

Figure~\ref{fig_2} illustrates the energy dependence of the electrical conductivity in a spin-1 chiral fermion system at absolute zero temperature. A distinct cusp structure is observed in the conductivity spectrum. We define the energy at which the cusp appears as the characteristic energy \( \epsilon_c \).
Importantly, this cusp does not occur precisely at the band crossing point \( \epsilon = 0 \), but rather slightly deviates from it. The absolute value \( |\epsilon_c| \) increases with increasing impurity strength \( W \) and decreasing curvature parameter \( |c| \). This behavior can be understood in terms of a band-resolved analysis, as explained below. 

As shown in Fig.~\ref{fig_3}(d), the intraband contribution from the trivial band drops rapidly below $\epsilon_c$, while that from the Dirac bands rises sharply above $\epsilon_c$, forming the cusp.
In this context, the cusp structure arises as a distinctive feature stemming from the multiband-crossing nature of the system. 
Figure~\ref{fig_3} shows that impurity scattering shifts the cusp energy $\epsilon_c$ away from the Dirac point. 
The shift of the cusp energy is related to the disorder-induced
renormalization encoded in the SCBA self-energy.
The real part of the SCBA self-energy renormalizes the quasiparticle energies and hence changes the effective position of the spectral weight relative to the bare Fermi energy.
At fixed carrier density, this effect may be viewed as a disorder-induced renormalization of the chemical potential. 
The imaginary part, on the other hand, gives the disorder-induced broadening. Since the nearly flat trivial band has a large low-energy spectral weight, both effects are enhanced near $\epsilon=0$ and shift the balance point between the Dirac-band and trivial-band conductivity contributions. 
This balance point determines the cusp energy $\epsilon_c$. For the positive curvature $c=0.1$ in Fig.~\ref{fig_3}, increasing $W$
therefore shifts the cusp to the positive-energy side.
As \( W \) increases, not only does \( |\epsilon_c| \) grow due to enhanced scattering, but both intraband and interband conductivity contributions also become smoother as functions of energy. This results in a more gradual cusp shape.

This behavior is closely related to the flat trivial band at \( \epsilon = 0 \), for which the density of states is singular in the clean limit. In the presence of impurities, this sharp spectral structure is broadened, and the SCBA is particularly suitable for describing this effect~\cite{Kikuchi2025} because it self-consistently incorporates the disorder-induced broadening of the density of states.

The value of \( \epsilon_c \) is determined not only by the impurity parameter \( W \) but also by the curvature parameter \( c \). 
Figure \ref{fig_4} presents the energy dependence of the conductivity for various values of \( c \), with the impurity strength fixed, emphasizing the distinct contributions from different scattering channels. 
In the special case of \( c = 0 \), the intraband contribution from the trivial band becomes exactly zero, and no cusp appears in the conductivity spectrum. 
For \( c \neq 0 \), however, the intraband contribution from the trivial band becomes finite and plays a key role in the emergence of the cusp structure. 
As \( |c| \) decreases, \( |\epsilon_c| \) increases. This trend can be attributed to the enhancement of the density of states near the Dirac point at smaller \( |c| \), rendering the system more susceptible to impurity scattering. 
Consequently, smaller \( |c| \) results in a smoother energy dependence of the conductivity, leading to a more gradual cusp shape. 

Furthermore, it is worth emphasizing that, as shown in Figs.~\ref{fig_4}(d)--(f), the intraband contribution from the Dirac bands is strongly suppressed over a wide low-energy range: (d) \( -0.22\, q_0 \hbar v \lesssim \epsilon \lesssim 0.22\, q_0 \hbar v \), (e) \( -0.28\, q_0 \hbar v \lesssim \epsilon \lesssim 0.20\, q_0 \hbar v \), and (f) \( -0.40\, q_0 \hbar v \lesssim \epsilon \lesssim 0.08\, q_0 \hbar v \). Such a broad suppression does not occur in systems consisting only of Dirac bands, such as Weyl fermion systems~\cite{0minato2014}. This feature can therefore be interpreted as a consequence of the hybridization between the Dirac bands and the trivial band induced by impurity scattering. The resulting reduction of the Dirac-band intraband contribution for \( \epsilon > 0 \) is one of the key factors responsible for the emergence of a sharp cusp structure.

In summary, these results demonstrate that the cusp structure is a prominent manifestation of the multifold band structure inherent to spin-1 chiral fermion systems. Physically, this cusp is formed by the competition between two transport channels with opposite energy dependencies: the intraband conduction within the Dirac bands and that within the trivial band of finite curvature. Impurity scattering plays a decisive role by broadening the energy levels near the Dirac point, which mixes the contributions from both channels. Consequently, the scattering strength $W$ and the trivial band curvature $c$ sensitively control the cusp's position $\epsilon_c$ and its corresponding conductivity $\sigma_c$. This physical picture, where impurity-induced level broadening is key to the formation of the cusp, underscores that the SCBA, which self-consistently incorporates this effect, is an essential framework for theoretically describing this phenomenon.

\section{Conclusion}\label{sec5}

In this study, we have analyzed the electrical conductivity of spin-1 chiral fermion systems with disordered potentials using the self-consistent Born approximation (SCBA) combined with vertex corrections. 
Building on previous studies of the energy dependence of the electrical conductivity in spin-1 chiral fermion systems, we have focused on how the band-resolved intraband and interband contributions depend on impurity scattering strength and the curvature of the trivial band.
The energy \( \epsilon_c \) at which the cusp emerges is determined by the strength of impurity scattering.  
As the scattering becomes stronger, both \( |\epsilon_c| \) and the corresponding conductivity \( \sigma_c \) increase.  
Moreover, when the curvature of the trivial band is small, the system becomes more susceptible to scattering, leading to further increases in \( |\epsilon_c| \) and \( \sigma_c \).
This unconventional energy dependence of the conductivity arises from the crossing of multiple energy bands protected by crystal symmetry. These findings serve as a basis for future investigations into novel transport phenomena in topological semimetals.

\backmatter

\bmhead{Acknowledgements}
This work was supported by JSPS KAKENHI (Grant Nos.\ JP25KJ1427, JP25K07224, and JP24H00853).




\bibliography{sn-bibliography}

\end{document}